\documentclass[12pt]{article}
\usepackage[margin=1.25in]{geometry}
\usepackage{graphicx}

\newcommand\pubnumber{SLAC--PUB--15966}
\newcommand\pubdate{May 2014}

\def\SLAC{SLAC,
    Stanford University, Menlo Park, California 94025 USA}
\def\doeack{\footnote{Work supported by the US Department of Energy,
                     contract DE--AC02--76SF00515.}}

\def\Title#1{\begin{center} {\Large #1 } \end{center}}
\def\Author#1{\begin{center}{ \sc #1} \end{center}}
\def\Address#1{\begin{center}{ \it #1} \end{center}}

\def\submit#1{\begin{center}Submitted to {\sl #1} \end{center}}
\newcommand\pubblock{\rightline{\begin{tabular}{l} \pubnumber\\
         \pubdate \end{tabular}}}
\newenvironment{Abstract}{\begin{quotation} \begin{center}
                       ABSTRACT
     \end{center}\bigskip  }{\end{quotation}}

\def\submit#1{\begin{center}Submitted to {\sl #1} \end{center}}
\def\Acknowledgements{\bigskip  \bigskip \begin{center} \begin{large}
             \bf ACKNOWLEDGEMENTS \end{large}\end{center}}

\def\beq{\begin{equation}}
\def\eeq#1{\label{#1}\end{equation}}
\def\eeqn{\end{equation}}

\newenvironment{Eqnarray}%
   {\arraycolsep 0.14em\begin{eqnarray}}{\end{eqnarray}}
\def\beqa{\begin{Eqnarray}}
\def\eeqa#1{\label{#1}\end{Eqnarray}}
\def\eeqan{\end{Eqnarray}}

\def\leqn#1{(\ref{#1})}

\let\bar=\overbar

\def\VEV#1{\left\langle{ #1} \right\rangle}

\def\ket#1{\left| {#1} \right\rangle}

\def\lsim{\mathrel{\raise.3ex\hbox{$<$\kern-.75em\lower1ex\hbox{$\sim$}}}}
\def\gsim{\mathrel{\raise.3ex\hbox{$>$\kern-.75em\lower1ex\hbox{$\sim$}}}}

\def\tr{{\mbox{\rm tr}}}
\def\half{\frac{1}{2}}
\def\thalf{\frac{3}{2}}
\def\third{\frac{1}{3}}
\def\tthird{\frac{2}{3}}

\def\del{\partial}
\def\Dslash{\not{\hbox{\kern-4pt $D$}}}
\def\dslash{\not{\hbox{\kern-2pt $\del$}}}

\def\ee{e^+e^-}

\def\msb{{\bar{\scriptsize M \kern -1pt S}}}

\def\drb{{\bar{\scriptsize D \kern -1pt R}}}

\makeatletter
\def\section{\@startsection{section}{0}{\z@}{5.5ex plus .5ex minus
 1.5ex}{2.3ex plus .2ex}{\large\bf}}
\def\subsection{\@startsection{subsection}{1}{\z@}{3.5ex plus .5ex minus
 1.5ex}{1.3ex plus .2ex}{\normalsize\bf}}
\def\subsubsection{\@startsection{subsubsection}{2}{\z@}{-3.5ex plus
-1ex minus  -.2ex}{2.3ex plus .2ex}{\normalsize\sl}}

\renewcommand{\@makecaption}[2]{%
   \vskip 10pt
   \setbox\@tempboxa\hbox{\small #1: #2}
   \ifdim \wd\@tempboxa >\hsize     
       \small #1: #2\par          
     \else                        
       \hbox to\hsize{\hfil\box\@tempboxa\hfil}
   \fi}

 \def\citenum#1{{\def\@cite##1##2{##1}\cite{#1}}}
 
\newcount\@tempcntc
\def\@citex[#1]#2{\if@filesw\immediate\write\@auxout{\string\citation{#2}}\fi
  \@tempcnta\z@\@tempcntb\m@ne\def\@citea{}\@cite{\@for\@citeb:=#2\do
    {\@ifundefined
       {b@\@citeb}{\@citeo\@tempcntb\m@ne\@citea\def\@citea{,}{\bf ?}\@warning
       {Citation `\@citeb' on page \thepage \space undefined}}%
    {\setbox\z@\hbox{\global\@tempcntc0\csname b@\@citeb\endcsname\relax}%
     \ifnum\@tempcntc=\z@ \@citeo\@tempcntb\m@ne
       \@citea\def\@citea{,}\hbox{\csname b@\@citeb\endcsname}%
     \else
      \advance\@tempcntb\@ne
      \ifnum\@tempcntb=\@tempcntc
      \else\advance\@tempcntb\m@ne\@citeo
      \@tempcnta\@tempcntc\@tempcntb\@tempcntc\fi\fi}}\@citeo}{#1}}
\def\@citeo{\ifnum\@tempcnta>\@tempcntb\else\@citea\def\@citea{,}%
  \ifnum\@tempcnta=\@tempcntb\the\@tempcnta\else
  {\advance\@tempcnta\@ne\ifnum\@tempcnta=\@tempcntb \else\def\@citea{--}\fi
    \advance\@tempcnta\m@ne\the\@tempcnta\@citea\the\@tempcntb}\fi\fi}
\makeatother

\begin{document}
\begin{titlepage}
\pubblock

\vfill
\Title{Ken Wilson:  Solving the Strong Interactions}
\vfill
\Author{Michael E. Peskin\doeack}
\Address{\SLAC}
\vfill
\begin{Abstract}
Ken Wilson's ideas on the renormalization group were shaped by his
attempts to build a theory of the strong interactions based on the
concepts of quantum field theory. I describe the
development
of his ideas by reviewing four of Wilson's most important papers.
\end{Abstract}
\vfill
\submit{the  Journal of Statistical Physics \\  Special Issue in
  Memory of K. G. Wilson}
\vfill

\newpage
\tableofcontents
\end{titlepage}

\def\thefootnote{\fnsymbol{footnote}}
\setcounter{footnote}{0}

\section{Introduction}

Ken Wilson is best known for his contributions to statistical
mechanics.  His breakthroughs in this field, including the 
computation of critical exponents and the solution of the Kondo
problem, have had wide influence.  Wilson began his career, however,
as an elementary particle physicist.  His ambition was to ``solve the
strong interactions'', that is, to find a predictive theory of the
subnuclear
strong interactions.   The ideas that he developed profoundly
influenced
our understanding of that problem, just as they provided tools and
insights for problems in statistical physics.  

 In this article, I will 
review the development of Wilson's ideas on the strong interactions
through a review of four of 
his most important papers~\cite{KGWone,KGWtwo,KGWthree,KGWfour}. 
I recommend these papers to all students of theoretical  physics. 
All four read like explorations of realms previously unknown.
They give insight into the  problems 
Wilson sought to address with his initial concepts of the
renormalization group.   And, both for 
particle physicists and for condensed matter physicists, they 
illustrate how issues in each domain
gave insight into the other.

\section{The Fixed Source Problem}

Quantum Field Theory (QFT) had some of its greatest successes in the
late 1940's, with the development of Quantum Electrodynamics and the 
successful explanation of the electron magnetic
moment and the Lamb shift.  The resulting euphoria led to the idea
that QFT could be used to build a theory of the strong interactions
based on a Lagrangian for pion-nucleon interactions.
The new technology of Feynman diagrams assisted calculation (as 
Feynman recalled memorably in his Nobel Prize lecture~\cite{FeynmanNobel}). However, it
did not produce a better understanding of the nuclear forces.  Pion exchange did not lead in any clear way
to the observed phenomenology of nucleon-nucleon scattering.  It could
not account for the nucleon and meson resonances that began to be 
discovered. 

 Most importantly, the theory had few concrete
predictions.  It was stymied by the fact that the strong interactions
are strong, while the methods developed for Quantum Electrodynamics 
relied on weak-coupling perturbation theory.  Strong coupling in QFT
implies that states with an arbitrarily large number of interacting
quanta play an essential role.  Feynman diagrams, which introduce
additional quanta one by one, cannot easily give insight into this
strong-coupling limit.

\begin{figure}[t]
\begin{center}
\includegraphics[width=0.7\hsize]{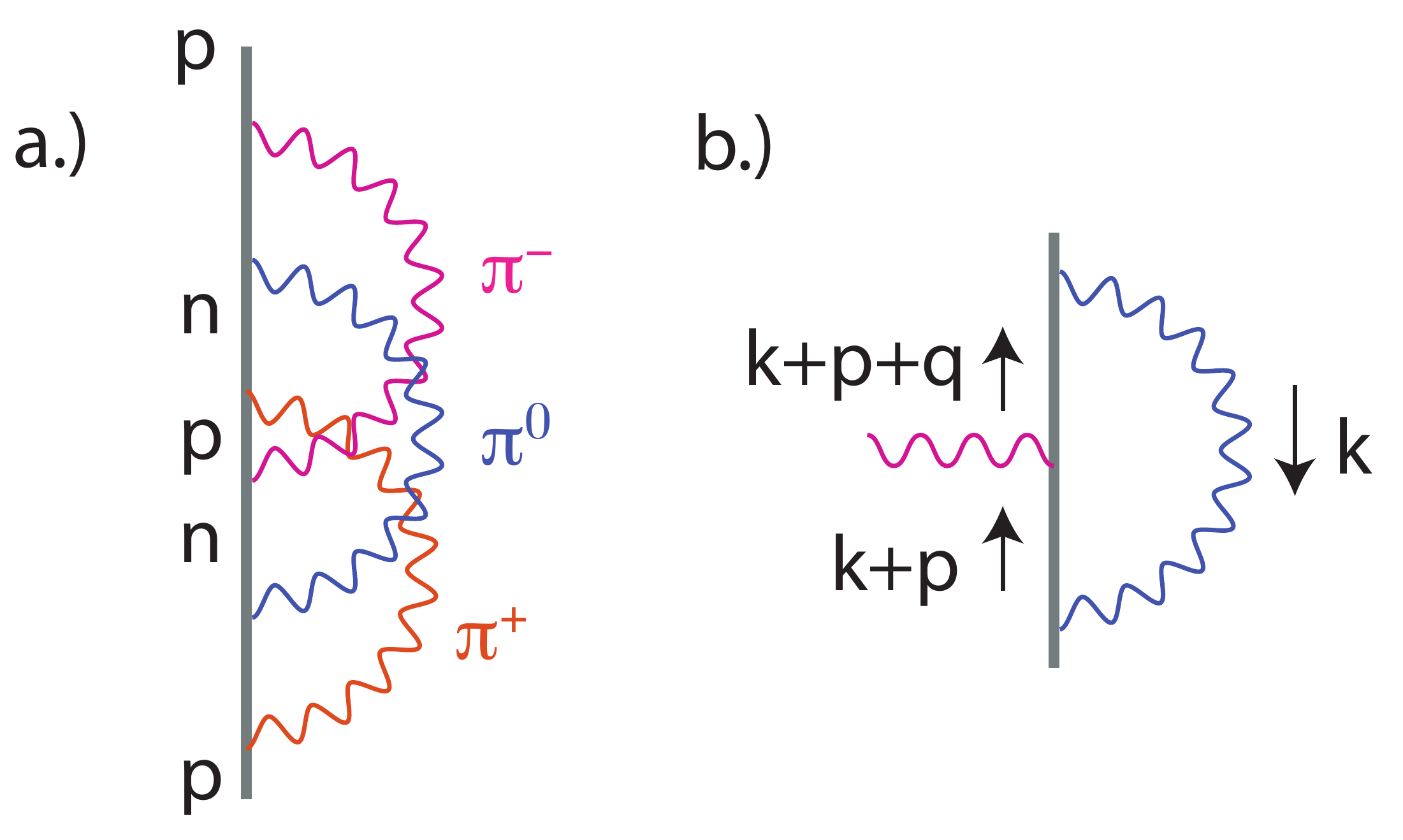}
\end{center}
\caption{Feynman diagrams for the fixed source problem: (a) a diagram
  illustrating the complexity of the problem; (b) a divergent one-loop diagram.}
 \label{fig:fixedsource}
\end{figure}

A relatively simple problem that encapsulated the difficulties of QFT
is
the fixed source problem.   This is the problem of a static or
infinitely  heavy nucleon with two states
\beq
   \ket{p} \qquad \ket{n}
\eeq{pnstates}
interacting at its location $\vec x = 0$ with a pion field 
\beq
   \pi^a (x) =   (\pi^+(x), \pi^0(x), \pi^-(x)) \ .
\eeq{pionfield}
T. D. Lee showed that a truncated version of this model, with only the  
$\pi^+$ field, could be solved exactly~\cite{Leemodel}.   However, the 
full problem allows complex intermediate states with many virtual
pions, 
shown in Fig.~\ref{fig:fixedsource}(a) as loops coupling to
the nucleon.  Further, each loop has an
ultraviolet
divergence.  The diagram shown in Fig.~\ref{fig:fixedsource}(b)
has the value
\beq
    -i   g^3 \int {d^4k\over (2\pi)^4} {1\over (k^0 + p^0) (k^0 +
      p^0+q^0) (k^2 - m_\pi^2)} \  ,
\eeq{diagramval}
and is logarithmically divergent.  So there is no
clear way even to compute the first loop sensibly, much less 
to limit the number of loops relevant to the final answer. 

By the mid-1950's, the search for the theory of the strong
interactions
had turned away from QFT to other methods, essentially 
phenomenological techniques such as 
dispersion relations and more fundamental proposals based on the 
analytic properties of scattering amplitudes.  Geoffrey Chew proposed that
there
was a unique analytic S-matrix that could be discovered by deep 
analysis.  As late as 1968, he stated: 
\begin{quote}
``There exists at present no
mechanical
 framework consistent with both quantum and relativistic principles.
 The chief candidate is local Lagrangian field theory, but countless
 theoretical studies have suggested insuperable pathologies in the 
concept of interaction between fields at a point of
space-time.''~\cite{Chew}
\end{quote}

Even for those who tried to build up the theory of strong interactions 
from symmetry principles, the infinities of quantum field theory posed
a barrier to taking this theory completely literally.   For example,  Murray
Gell-Mann's paper that introduced the method of current algebra---one
of the most important theoretical methods used in particle physics in
the 1960's---
includes the following statement: 
\begin{quote}
``... we use the method of 
abstraction from a Lagrangian field theory model. In other words, we 
construct a mathematical theory of strongly interacting particles,
which 
may or may not have anything to do with reality, find suitable
algebraic 
relations that hold in the model, postulate their validity, and then 
throw away the model. We may compare this process to a method 
sometimes employed in French cuisine: a piece of pheasant is cooked 
between two slices of veal, which are then discarded.''~\cite{GellManncurrent}
\end{quote}

Ken Wilson was Gell-Mann's student at Caltech from 1957 to 1961.  He
chose the fixed source problem described above as the topic of his
thesis research.  In his thesis, he threw at this problem the full
arsenal of mathematical methods developed in the 1950's, with 
minimal success.  This investigation proved Wilson's skills and promise,
but it not make much headway toward the solution.

Many first-rank theorists find their thesis problem overreaching and 
frustratingly difficult.  Usually, the solution is to pick another
problem that yields more easily to their talents.   This was not
Wilson's style.  He would continue to struggle with the fixed-source
problem for many more years.

\section{Momentum Slicing}

Wilson's first  published work on the fixed-source problem did not
appear until 1965.  It is the first of the four papers that are the
subject of this review:
\begin{quote}
``Model Hamiltonians for Local Quantum Field Theory'',
Phys. Rev. {\bf 140}, B445 (1965) \cite{KGWone}
\end{quote}
This paper had no immediate impact, because it enunciated a point of
view that ran against the main current of theoretical particle physics 
in the 1960's.  Wilson's boldly stated attitude is that 
 there is no mysticism about QFT. The way to understand
its  issues to reduce problems in QFT to 
ordinary quantum-mechanical problems that can be solved
by the standard methods of atomic physics.  As Wilson writes in this 
paper:
\begin{quote}
``The Hamiltonian formulation of quantum mechanics has been essentially
abandoned in the investigations of the interactions of $\pi$  mesons,
nucleons,
 and strange particles. This is a pity. The Hamiltonian approach has 
several advantages over the kind of approach (using dispersion
relations) 
presently in use.  One advantage is that all properties of a system 
are uniquely determined ... A second advantage is the existence of
many 
approximation schemes ...  A third advantage is that one can 
often analyze a Hamiltonian intuitively.''~\cite{KGWone}
\end{quote}
For the neglect of QFT, Wilson blamed the problem of infinities.  To 
rectify this, what was needed was a direct assault on that 
problem.

The infinities of QFT arise from the fact that the quantum excitations 
represented by the legs of Feynman diagrams may have any momenta and, 
in particular, momenta taking  arbitrarily high values.  Wilson's
approach
to the infinities was to lay out these momenta in an orderly set of
regions 
that could be analyzed one by one.   He called this concept ``momentum
slicing''. 

\begin{figure}[t]
\begin{center}
\includegraphics[width=0.6\hsize]{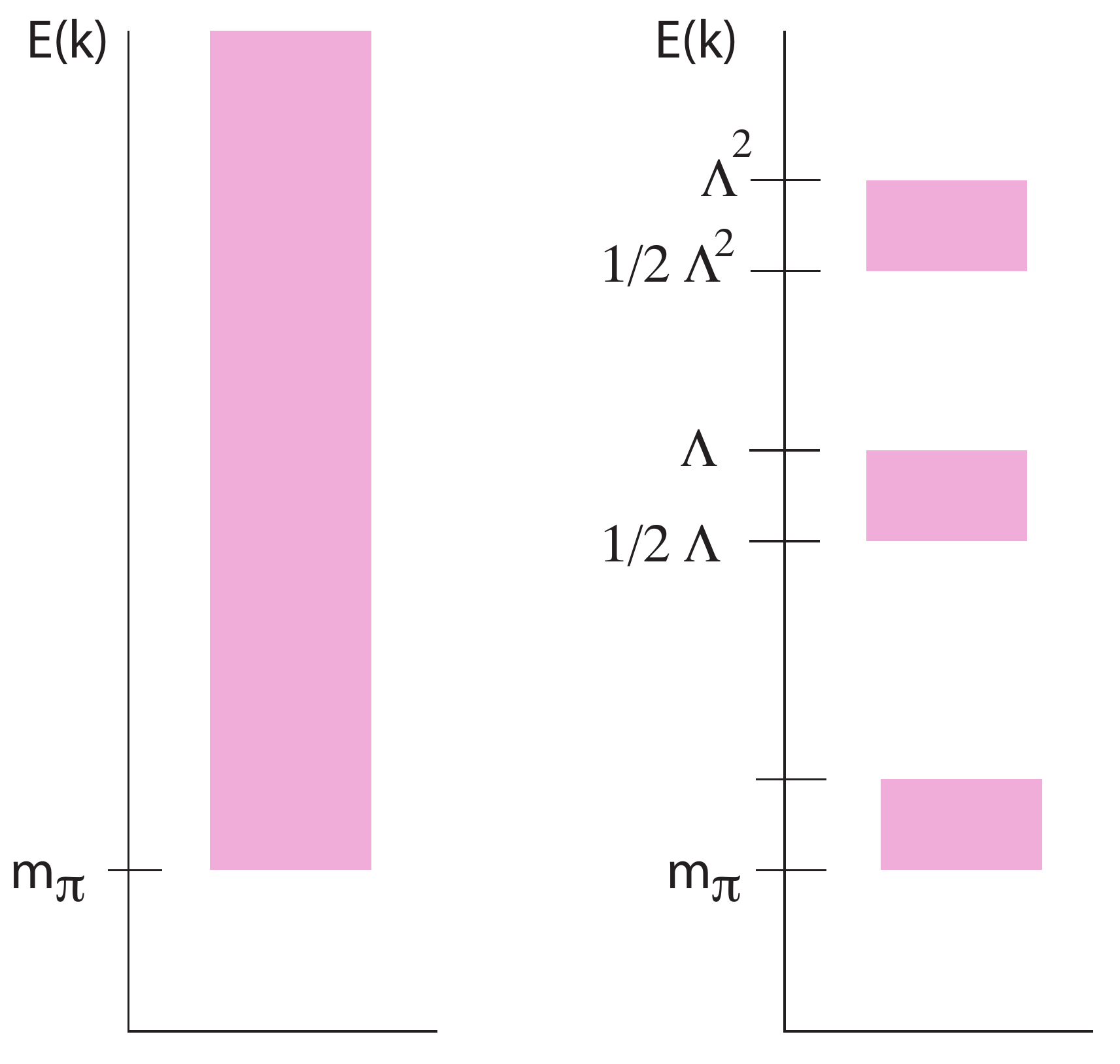}
\end{center}
\caption{The slicing of momentum space used by Wilson in
  \cite{KGWone}.  The vertical axis gives the energies associated with
  the selected momentum states, on  a logarithmic scale.}
\label{fig:slicing}
\end{figure}

 In \cite{KGWone}, the full momentum space available to
pions in the fixed source problem  is replaced by
the 
set of intervals
\beq
  0 < |k| < m_\pi \ , \quad \half \Lambda < |k| < \Lambda\ , \quad
  \half \Lambda^2 < |k| < \Lambda^2 \ , \quad \ldots , \quad  
\half \Lambda^n < |k| < \Lambda^n \ ,
\eeq{firstslices}
The slicing of momentum space is illustrated in
Fig.~\ref{fig:slicing}. 
This severe reduction of the allowed phase space not only  sharpens the problem
of infinities but also reverses the standard viewpoint.   At first
sight, it is the low-momentum degrees of freedom that are the most
important for the physics of pion-nucleon interactions encoded by the 
fixed-source problem.  The appearance of high momenta is an intrusion that needs
to be controlled.  However, if the system defined by
\leqn{firstslices} is studied using the standard approximation schemes
of quantum mechanics, the opposite is true.  The most important terms
in the Hamiltonian are those in the highest momentum interval.  This
part of the Hamiltonian must be diagonalized before lower-momentum 
intervals can be studied.

Standard ideas of quantum-mechanical perturbation theory dictate how
this diagonalization should be done.  The problem of pions in the
highest momentum interval should be solved first, and the ground state
of this system found.  This ground state configuration of the pion modes
with momenta of order $\Lambda^n$ can then be used as the starting
point for an analysis of the pion modes with momenta of order
$\Lambda^{n-1}$.  It is only at the end of this process that momenta 
of the order of $m_\pi$ come into play.

The diagonalization of the Hamiltonian is then naturally structured as
an iteration.   The modes at  $|k| \sim \Lambda^n$ primarily affect
the modes at $|k| \sim \Lambda^{n-1}$ by modifying their coupling to
the nucleon.   This gives  a recursion equation
\beq
           g_{n-1}  =   f(g_n) 
\eeq{recurse}
The effect of modes at large momentum scales on the pion modes at
$|k| \sim m_\pi$ is then encapsulated in the evolution of the coupling
constant from scale to scale that results from this evolution. At each
stage of the evolution, the higher momenta are said to be ``integrated
out'' and are removed from Hamiltonian.   This
is the essence of Wilson's concept of the renormalization group.

The restriction of momenta to the domains \leqn{firstslices} is of
course an extreme truncation of the original problem.   To solve
the fixed source problem quantitatively by momentum slicing,
 it is necessary to consider regions that are continguous, without
 gaps, and to integrate out each high-momentum region down to its
boundary in an accurate way.  In \cite{KGWformal}, Wilson addressed
this question as a matter of principle, proving that the method led to a
solution to the fixed source problem with all infinities eliminated.

However, Wilson did not stop there.  It happens that the fixed source
problem is related to a puzzle that appeared in the theory of
magnetism, the Kondo problem~\cite{Kondo}.   This is the problem of a fixed
magnetic impurity coupling to a free gas of electrons.   For
ferromagnetic coupling, the impurity behaves as a weakly coupled
free spin interacting with the electrons. But for antiferromagnetic 
coupling, however weak in the underlying theory, the ground state 
contains a strong binding of the impurity to an electron that
essentially quenches its magnetism. By introducing additional
operators and corresponding couplings that transform under a
multiparameter recursion equation, Wilson was able to integrate out
shells in the electron momentum sufficiently accurately to identify
the transition energy from weak to strong coupling.   The full story
of this calculation is outside the scope of this review, but it is 
described lucidly in \cite{WilsonKondo}.   Wilson's calculation of the 
strong-coupling scale in this model
\beq
  {  T_K\over 4\pi T_0} = 0.1032 \pm 0.0005
\eeq{Kondorelation}
was later verified by an exact solution of the Kondo problem, using 
Bethe's ansatz, by Andrei and Lowenstein~\cite{AL}.

\section{The Operator Product Expansion}

\begin{figure}[t]
\begin{center}
\includegraphics[width=0.2\hsize]{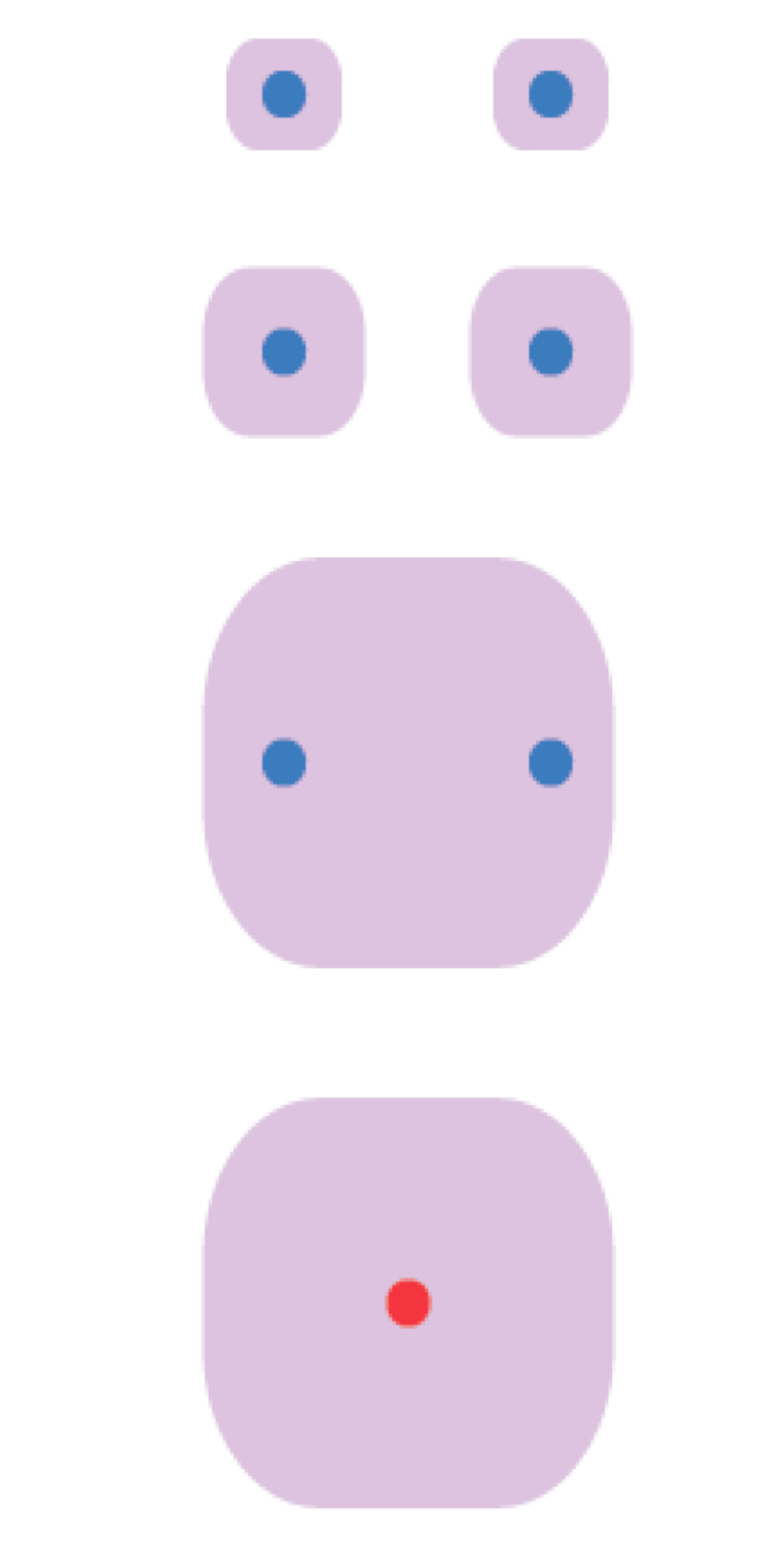}
\end{center}
\caption{Generation of a composite operator by integrating out
  high-momentum degrees of freedom.}
\label{fig:merge}
\end{figure}

The idea of integration out has a more general consequence for the
description of operators in QFT.   Integrating out momentum modes 
with $|k| \sim \Lambda$ corresponds to the solution of the quantum 
theory for point separatons of order $|x-y|  \sim \pi/\Lambda$.
Local operators placed more closely together that this distance
cannot be considered separately after integration out of this 
mometum shell.  They  must 
merge to become single operators located at some intermediate 
point, as shown in Fig.~\ref{fig:merge}.

Formalizing this intuition gives Wilson's Operator Product Expansion.  The
idea of the operator product expansion appeared fully formed in the 
literature in the second paper of this review
\begin{quote}
``Non-Lagrangian Models of Current Algebra'', Phys. Rev. {\bf 179}, 1499
(1969).~\cite{KGWtwo}
\end{quote}
The concept is expressed by the statement that all expectation values
of a pair of QFT operators located at points $x$ and $y$ together with 
operators located at points $z$ far from $x,y$ can be computed by 
the replacement
\beq
 {\cal O}_A(x) {\cal O}_B(y) = \sum_C  \  {\cal C}_{ABC}(x-y) {\cal
   O}_C(y)  \ ,
\eeq{OPE}
where the sum over $C$ runs over all operators in the QFT with
appropriate 
quantum numbers, and $ {\cal C}_{ABC}(x-y) $ is a c-number coefficient
function.

The relation is especially simple in theories in which the QFT
dynamics at distances smaller than $|x-y|$ is independent of any
intrinsic length scale.  Then, for scalar operators
\beq 
      {\cal C}_{ABC}(x-y) =    {c_{ABC}\over |x-y|^{d_A +d_B - d_C} }
      \ ,
\eeq{simpleOPE}
where $c_{ABC}$ and $d_A$, $d_B$, $d_C$ are numbers.   For operators
with spin, $c_{ABC}$ is replaced by a number times an appropriate
Lorentz structure. 

 The quantities $d_i$ are called the dimensions of the
 operators
and reflect the scaling of operator matrix elements with changes of distance
scale.  In Wilson's original conception, these dimensions were
integers, as in free field theory.   An
anonymous referee (now known to be Arthur Wightman~\cite{Sch}) called
 Wilson's attention to the Thirring model, an exactly solved model in
 (1+1) 
dimensions, and prompted Wilson to make a serious study of this model.
 In the Thirring model, operator dimensions can be arbitrary real
 numbers.
In \cite{KGWtwo}, the idea that the
values of
operator  dimensions could have a nontrivial influence on physical
phenomena was presented for the first time.

The paper \cite{KGWtwo} applied these ideas to one of the most
important problems being considered at that time, the nature of 
products of currents.    Such products appear in the structure of the 
weak interactions, in analyses of the properties of pions and kaons
using the methods introduced in \cite{GellManncurrent}, and in the 
analysis of deep inelastic electron scattering.   In 1967, the results
of
the SLAC-MIT electron scattering experiment and their interpretation 
by Bjorken~\cite{BJ} and Feynman~\cite{partonmodel} pointed to 
models of the structure of the proton with free-field behavior at 
short distances for the proton constituents.  Theoretical analysis of
these experiments required the high-momentum asymptotic behavior of a
pair of electromagnetic currents.

One of the properties of the  Thirring model that was striking to
Wilson in this context is that the current algebra of the model
remains unchanged as the operator spectrum of the model is distorted
by the effects of strong interactions.  In our (3+1)-dimensional
world, conserved currents would remain operators of dimension $d =3$
while the dimensions of other operators
 would shift. Typically, results derived from
Gell-Mann's current algebra for strong interaction
matrix 
elements depend not only on the algebra but also on the behavior of 
short-distance limits.  The operator product expansion gave a
systematic
way to analyze this issue.

The result of the paper that seems most striking from our modern point
of view is the explanation Wilson gives for  the $\Delta I = \half$ rule, the 
fact that $\Delta I = \half$ weak decays of $K$ mesons and strange baryons, for example
$K^0\to \pi^+\pi^-$, go more than 100 times faster than $\Delta I =
\thalf$  decays such as 
$K^+\to \pi^+\pi^0$.  Wilson suggested that different operators,
${\cal O}_C$
in \leqn{OPE}, in the
product of $W$ boson currents
contributed to these two amplitudes, and that the  difference in the
amplitudes  arises from the different factors
\beq
                          (m_K/m_W)^{6-d_C}  \ . 
\eeq{ratdC}
in their operator product coefficients.    This was the first  suggestion of a qualitative
 effect on physics caused  by dynamically-generated differences in 
operator dimensions.  In 1974, Gaillard and Lee analyzed the operator
 product of $W$ boson currents in the gauge theory of strong
 interactions 
QCD, to be described below, and showed that this effect does account for a large part, if
not all, of the $\Delta I = \half$ enhancement~\cite{GaillL}.

\section{Scale Invariance at Short Distances}

Wilson discussed the results of the paper \cite{KGWtwo}  in the context
of a vision for the structure of a QFT description of strong
interactions.
The infinities of the theory would be tamed by the principle that the
recursion described in Section~3 would have converged to a  set of
couplings that did not change with scale.  Wilson describes a 
``skeleton theory'' for strong interactions that is exactly scale 
invariant.  To build a realistic model with nonzero hadron masses,
this theory would be perturbated by mass terms or other operators 
with dimensionful coefficients.   Wilson notes the idea of Kastrup~\cite{Kastrup}
and Mack~\cite{Mack} that these terms might arise from the spontaneous
breaking of a scale symmetry. 

I  have already pointed out  that the idea that the strong
interactions are described
at short distances by a scale-invariant free field theory
was   very much in
the air at this time.    Both current algebra and the parton
description of deep inelastic scattering rested on this foundation.
However, it was recognized that this foundation could not be realized
in any interacting QFT model.  

Wilson's ideas cut through the haze 
surrounding this question.   They suggested a framework for a model
that could actually arise from QFT.  However, the exact form of that 
model was still obscure.  Wilson ends the paper \cite{KGWtwo} with the
statement:
\begin{quote}
 ``It is hard to imagine that one could have a complete formula ...
without  having a complete solution of the hadron
 skeleton theory.  The prospects for obtaining such a solution seem
dim at present.''~\cite{KGWtwo}
\end{quote}

\section{The Paper with Three Errors}

If all one knows about the underlying scale-invariant theory of strong 
interactions is that it arises from renormalization group recursion,
one can at least analyze the possibilities for the structure of such a theory
by examining the renormalization group equations more closely. 
Wilson presented such an analysis in the third paper reviewed here,
\begin{quote}
 ``Renormalization Group and Strong Interactions'', Phys. Rev.
 {\bf D3}, 1818 (1970)~\cite{KGWthree}.
\end{quote}

This paper concentrates on the case of one coupling constant evolving
according to a continuous  renormalization group equation.  In modern
notation, this equation is
\beq
              {d g(\mu)\over d \log \mu } = \beta(g(\mu)) \ ,
\eeq{beta}
where $\mu$ is a momentum scale,
 $g(\mu)$ is the dimensionless coupling constant of the strong 
interaction theory, and $\beta(g)$ describes its evolution with scale.
Wilson did not consider this equation familiar to his  auidience.
Rather,
much of the paper is devoted to deriving the equation in perturbative
QFT, beginning from the original treatment  of Gell-Mann and
Low~\cite{GML}.
Wilson gives  as his primary reference for the renormalization group
the textbook of Bogoliubov and Shirkov~\cite{BS}, though his
explanation is a highly processed version of the one found
there~\cite{tome}.

\begin{figure}[t]
\begin{center}
\includegraphics[width=0.5\hsize]{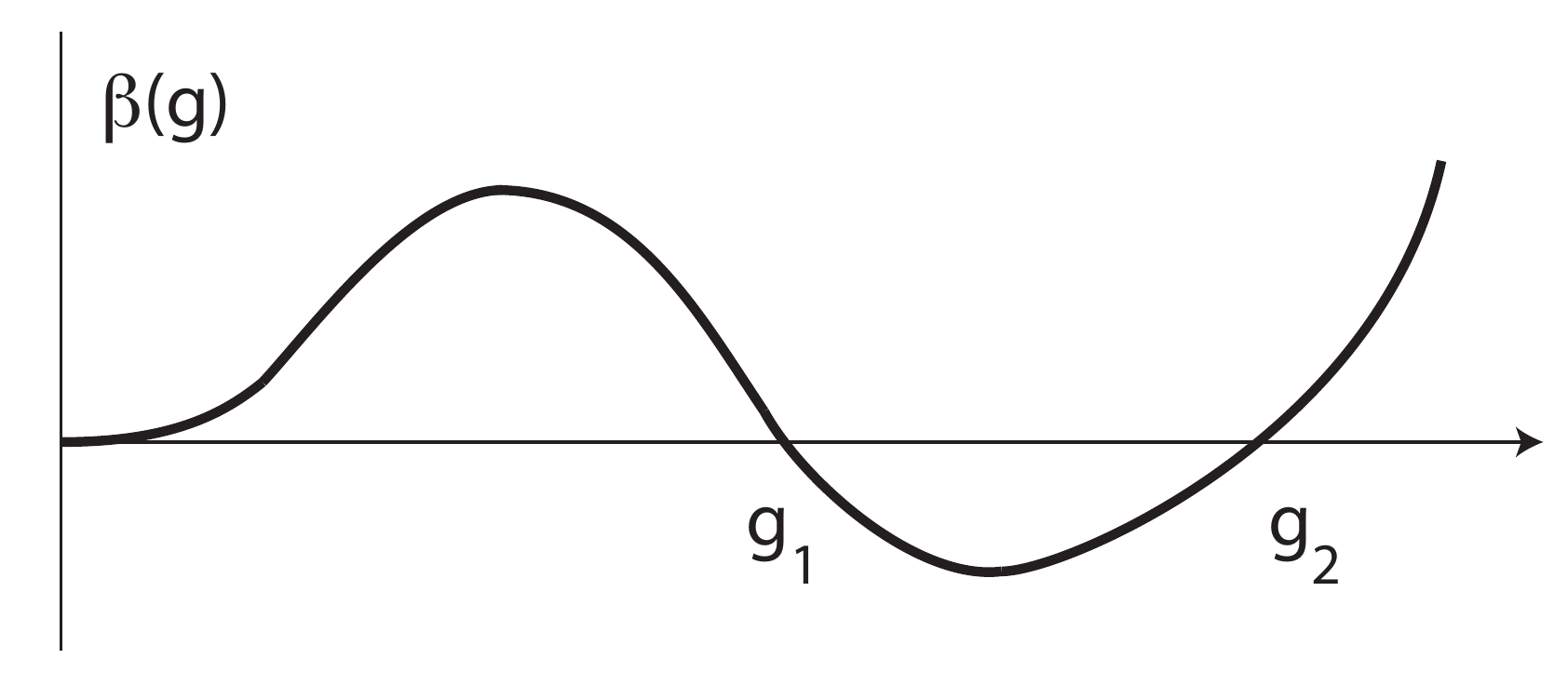}
\end{center}
\caption{An example of a renormalization group function $\beta(g)$.}
\label{fig:betaf}
\end{figure}

Wilson's approach to \leqn{beta} was to analyze it as the 
equation for a general dynamical system.   Possible asymptotic 
behaviors for a dynamic system include a fixed point and a limit
cycle.   Wilson described fixed-point solutions to \leqn{beta} with the example
of a $\beta$ function of the form shown in Fig.~\ref{fig:betaf}.  At 
momenta $M$ high enough that mass parameters could be ignored, the 
coupling constant $g$ would take some value $g_M$ on the horizontal 
axis.  For values  $g_M < g_1$, the value would then increase at 
larger scales, coming close asymptotically to the value 
$g_1$,  which would be a fixed point of the renormalization group.
If the value of $g_M$   were in the range $g_1 < g_M < g_2$. the value
of  $g(\mu)$ would decrease to the fixed point $g_1$.

An  alternative picture discussed by Wilson is one in which, at very high
momentum scales, the weak and electromagnetic couplings become as
large at the strong interaction coupling.  The high-momentum value of
$g$ is then determined by properties of this  unified theory.   At
lower scales, where the strong interactions can be treated in
isolation from the weak and electromagnetic interactions, $g$ takes on
the renormalization group evolution described by $\beta(g)$.   In that
case, the low-momentum behavior might be evolution to an
infrared-asymptotic
fixed point of the renormalization group, such as $g_2$.   The
observed
strong interaction coupling would be modified slightly from $g_2$ by 
effects of the mass terms.

Wilson analyzes one more possibility.   In principle, if there is more
than one  coupling constant, the high-energy asymptotic behavior of
the renormalization group equation might be a limit   cycle.  In that
case, the   values of the coupling constants would perpetually
oscillate, with a regular period in $\log\mu$.   This would be
directly 
observable as an oscillating behavior of the cross section for $\ee$ 
annihilation to hadrons.

This paper was eye-opening for many theorists at the time~\cite{Gross}.  It
introduced the idea of qualitative analysis of a QFT through
visualization of its renormalization group flows, an idea that is now
a standard method both in particle physics and in statistical
mechanics.

Still, Wilson is said to have refered to this work as ``the paper with
three errors''.   In hindsight, the omissions are easy to find.  Limit
cycles of the renormalization group have never played a role in
particle
physics (though examples do exist for discrete renormalization group
transformations~\cite{Glazek}).    The idea that the low energy value
of the strong interaction coupling constant would be an
infrared-stable fixed point of the renormalization group was also not
reallized.   This idea might still be relevant in the theory of the
top
quark mass~\cite{PRoss,Hill}.  Most importantly, though, Wilson assumed that
the
$\beta$ function must be positive in the low-$g$ region of
Fig.~\ref{fig:betaf},  where it is computed
in 
perturbation theory.  He writes that, negative  $\beta(g)$ ``violates
the K\"all\'en-Lehmann representation for the photon propagator''~\cite{KGWthree}. 
The last secret of the strong interactions was hidden here, as I will
explain in a moment.

\section{Statistical Mechanics and Quantum Field Theory}

It was in this same period that Wilson became involved with problems
of the theory of phase transitions.   Wilson's interaction with David
Mermin,
Michael Fisher, Ben Widom, and other statistical mechanics experts at
Cornell will be covered by other contributions to this volume.  
 The street ran both ways.  The
converse of the 
idea that statistical mechanics problems can be modelled by QFT is the
idea that QFT can be given a foundation by constructions taken
from statistical mechanics.  

From our previous discussion, the ingredients are all in place.   A
lattice with spacing $a$ in $d$ dimensions can represent the space that 
results when quantum states at large  momenta are integrated out down
to $|k| \sim \pi/a$.
Integration out potentially leads to a complicated Hamiltonian with
many nonzero operator coefficients.  However, most of these operators
have high dimension and so do not affect
physics at energy scales of the order of particle masses.    

The cleanest connection of this type is between lattice statistical
mechanics problems on a $d$-dimensional lattice and Euclidean QFT 
in $d$ dimensions.  It follows from the axioms of QFT~\cite{StrW} that operator
expectation values on Lorentzian spacetime can be analytically
continued
to a Euclidean spacetime with 
\beq
        x^2 =   (x^0)^2 + (x^1)^2 + (x^2)^2 + (x^3)^2 \ .
\eeq{Euclid}
Continuation to Euclidean space carries the time translation operator
\beq
    U(t) =    e^{-iHt} \to    T(t_E) =  e^{- L_Et_E }\ ,
\eeq{toSM}
where $L_E$ can be identified with the Lagrangian of the analytically continued problem.
The object on the right is the transfer matrix of a statistical
mechanics
problem.  Setting $t_E =a$ gives the evolution from one lattice
spacing to the next.   The complete partition function is
\beq
    Z =   \tr [   T(a)^N ] 
\eeq{partition}
for a lattice of length $Na$.

Wilson's student Ashok Suri worked out these connections in detail and
explained them in his 1969 Ph.D. thesis~\cite{Surithesis}.    That
thesis became a basic reference document for the developments to follow.

\section{Quantum Chromodynamics}

The missing piece in the story of the scale invariance of strong
interactions popped out in the spring of 1973 with the announcement by 
Politzer, Gross, and Wilczek that non-Abelian Yang-Mills theory is
asymptotically free~\cite{Politzer,GW}.  By this, I mean that the
renormalization group equation in this theory has a negative 
$\beta$ function at small values of $g$, causing the coupling
constant to run to zero for large momenta. In (3+1)-dimensions, 
non-Abelian gauge theories are absolutely exceptional in allowing 
negative values of the $\beta$ function~\cite{Zee,CG}.  The indefinite
metric spaces used in the quantization of gauge fields 
allow them to  slip through a crack in the argument that the $\beta$
function is always positive.

  A theory
with
asymptotic freedom would be asymptotically scale-invariant, and, in
fact,  
free, at short distances.   However, in certain observables, one could still find large
effects of operator dimensions, now scaling with powers of logarithms
of momenta
\beq
                          \biggl(  \log {m_K^2\over \Lambda^2} /
                          \log{m_W^2\over \Lambda^2}\biggr)^{\gamma_A
                            + \gamma_B - \gamma_C}
\eeq{rather}
rather than \leqn{ratdC}.    The combination of these two features in
the same package made this an ideal solution to all of the problems of
constructing a theory of strong interactions.
 Almost 
immediately, the idea of building interactions from gauge fields
 converged with other aspects of strong 
interaction phenomenology to pick out the Yang-Mills gauge group 
$SU(3)$, with quarks belonging to the fundamental {\bf 3}
representation~\cite{QCD}.    This theory was
named Quantum Chromodynamics (QCD).  Today,  a wealth of evidence
supports the claim that QCD is the fundamental description 
of the strong interactions.

QCD included a picture of strongly interacting particles as bound states of more
fundamental
spin $\half$ particles, quarks.   The quark model explained the mass spectrum 
and quantum numbers of the baryons and mesons.   It gave a
basis for current algebra and the structure of hadronic weak
interactions.
But, together with these successes came a puzzle:   The quark model 
required the electric charge assignments $+\tthird$ for the $u$ quark
and $-\third$ for the $d$ and $s$ quarks.  Yet, no particles with 
fractional electric charge had been seen in nature.  By 1973,
extensive
searches had been done, all with negative results~\cite{noquarks}.

\section{Lattice Gauge Theory}

Wilson had not been studying gauge theories, or    any weak-coupling
proposal
for the nature of the skeleton theory of strong interactions.  However, now a
question arose that he was uniquely positioned to answer:  How do
non-Abelian 
gauge theories behave when their coupling constants are taken to be strong?
The answer to this question was given in the final paper for this
review
\begin{quote}
 ``Confinement of Quarks'', Phys. Rev. {\bf D10}, 2445 (1974)~\cite{KGWfour}
\end{quote}

That answer turned out to be profound.  The following important
results were summarized in the introduction to this paper:
\begin{quote}
``A new mechanism which keeps quarks bound will be proposed in this
paper. The mechanism applies to gauge theories only.''

``By analogy to the solid-state situation one can think of the
transition from zero to nonzero photon mass as a change of phase.''

``... the strong-coupling expansion ... has the
same general structure as the relativistic string model of hadrons
...''~\cite{KGWfour}
\end{quote}

The correspondence between continuum QFT and lattice statistical 
mechanics was the crucial tool for this investigation.
By now Wilson had thoroughly assimilated the idea of lattice-regulated
QFT.  He writes:
\begin{quote}
``The model discussed in this paper is a gauge theory set up on a
four-dimensional Euclidean lattice. The inverse of the lattice spacing
serves as an ultraviolet cutoff. 
 The use of a Euclidean space ... instead of a Lorentz space is not a
 serious restriction.''
\end{quote}
Other members of our particle physics community took decades to get
used to this idea.  Today, gauge theory on a Euclidean lattice  is a proven numerical
tool for calculations in the low-momentum region of
QCD~\cite{LGTreview}.

\begin{figure}[t]
\begin{center}
\includegraphics[width=0.6\hsize]{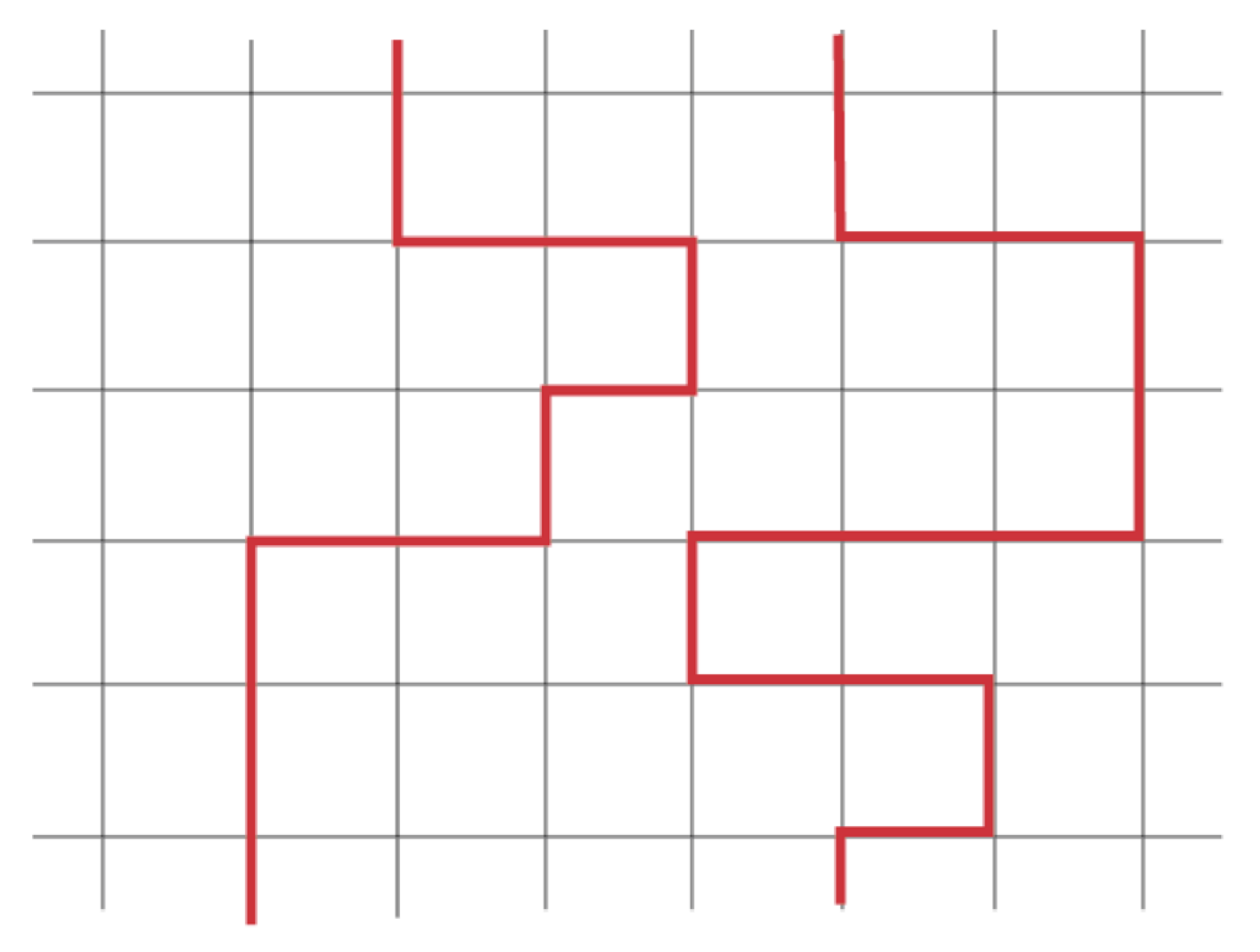}
\end{center}
\caption{Path of a heavy quark-antiquark pair  in a lattice
  space-time.  Euclidean time runs upward.}
\label{fig:globalpath}
\end{figure}

To begin the description of a local gauge symmetry on a lattice, we
might start from the description of a matter particle in a lattice
QFT.
  A path of a heavy particle in Euclidean spacetime
has the form shown in Fig.~\ref{fig:globalpath}.   A quantum
particle travelling on paths of this type can be realized  by a scalar
field transforming under a global symmetry, which we associate with
the particle number,  
\beq
     \phi_n \to  e^{i\alpha} \phi_n \ ,
\eeq{isalpha}
where $\alpha$ is a global parameter.
A discretized derivative of
the field can be defined as
\beq
   \Delta_\mu \phi_n = {1\over a} (\phi_{n + a\hat \mu} - \phi_n)
\eeq{discderiv}
A lattice QFT with the Lagrangian
\beq
    L_E =   \sum_n\ a^3\,\biggl[  {1\over 2T}  |\Delta_\mu\phi_n|^2 + \half m^2 |\phi_n|^2 \biggr]
\eeq{latticeL}
is invariant under the symmetry.  Expanding the partition function in
the parameter $1/T$ is a controlled expansion in the lattice-regulated
theory.  This expansion is analogous to the high-temperature expansion
of lattice statistical models.   It corresponds to a sum of graphs
with paths of the form shown in the figure.   A similar treatment can
be given for fermions on the lattice.  To implement this, Wilson
described
the integral over fermionic variables invented by
Berezin~\cite{Berezin}, still, at that time, quite unfamiliar as a
QFT tool. 

The generalization to a local gauge symmetry raises new issues.
The  symmetry transformation is now 
\beq
     \phi_n \to  e^{i\alpha_n} \phi_n \ ,
\eeq{isalphalocal}
   where $\alpha_n$ is independent at each lattice site.  The
   derivative
\leqn{discderiv} now no longer has a linear transformation law under
the symmetry group.

The remedy for this problem is to generalize the lattice derivative
with 
an additional element
\beq
   \Delta_\mu \phi_n = {1\over a} (\phi_{n + a\hat \mu} - U_{n+a\hat
     \mu,n}\phi_n) \ , 
\eeq{newdiscderiv}
where $U_{n_1,n_2}$, with $(n_1,n_2)$ neighboring lattice sites, has
the transformation
\beq
    U_{n_1,n_2} \to   e^{i\alpha_{n_1}} U_{n_1,n_2}  e^{-i\alpha_{n_2}}
\eeq{newlaw}
Minimally, $U_{n_1,n_2}$ can be taken to be a unitary matrix representing an
 element of the gauge group, with the identification  $U_{n_2,n_1} =
 U_{n_1,n_2}^\dagger$.  The statistical sum over $U_{n_1,n_2}$
 is an integral over the gauge group for each link of the lattice.

In terms of continuum variables, a quantity with the same
transformation law as $U_{n_1,n_2}$ is the exponential of the line
integral of the vector potential
\beq
       U_{n_1,n_2} \equiv    \exp\biggl[ i g \int_{n_2}^{n_1} dx^\mu
       A_\mu ]
\eeq{Uisline}
Making this identification and expanding for the case in which
$A_\mu(x)$ varies slowly over a lattice spacing, we find
\beq
       \Delta_\mu \phi =  (\del_\mu -i g A_\mu)\phi \ ,
\eeq{gcderiv}
the standard gauge-covariant derivative.   A gauge-invariant quantity
built purely from the $U_{n_1,n_2}$ is 
\beq
    \tr [V_{n,\mu,\nu} ] =    \tr[      U_{n,n+\hat\nu}
    U_{n+\hat\nu,n+\hat\mu+\hat\nu}
 U_{n+\hat\mu+\hat\nu,n+\hat\mu}
       U_{n+\hat\mu,n} ] 
\eeq{Usquare}
In the continuum, $V_{n,\mu,\nu}$ can be identified as the exponential
of a line integral around the  elementary square,
\beq
       V_{n,\mu\nu} \equiv   \exp\bigg[ i g \oint dx^\mu A_\mu \biggr]
       = \exp\biggl[ i g \int d^2s^{\mu\nu} F_{\mu\nu}\biggr]
\eeq{Vident}
I have written these formulae for the case of an Abelian gauge group,
but they go through with only minor modifications for a non-Abelian
gauge group.   The lattice QFT action
\beq
   L_E   =  \sum_{n,\mu,\nu}  {1\over g^2} \tr [ V_{n,\mu,\nu }+
     V_{n,\mu,\nu}^\dagger ] 
\eeq{GFLag}
then gives the usual gauge field action ${\cal L } = -{1\over 4}
(F_{\mu\nu})^2$.
Similar lattice QFTs were constructed by Wegner~\cite{Wegner}, for the 
case of discrete gauge symmetry, and by Polyakov~\cite{Polyak}.

The lattice Lagrangian \leqn{GFLag} has the amazing property of
possessing a straightforward expansion in powers of $1/g^2$.   One
simply needs to expand the exponential of \leqn{GFLag} in a power
series.
Each factor of $\tr[V_n]$ brings down four factors of $U_{n,n+\hat\mu}$,
which are then integrated over the gauge group.   This $1/g^2$
expansion realized one of Wilson's longstanding goals, to 
directly compute the structure of a QFT in the limit of strong coupling.

\begin{figure}[t]
\begin{center}
\includegraphics[width=0.6\hsize]{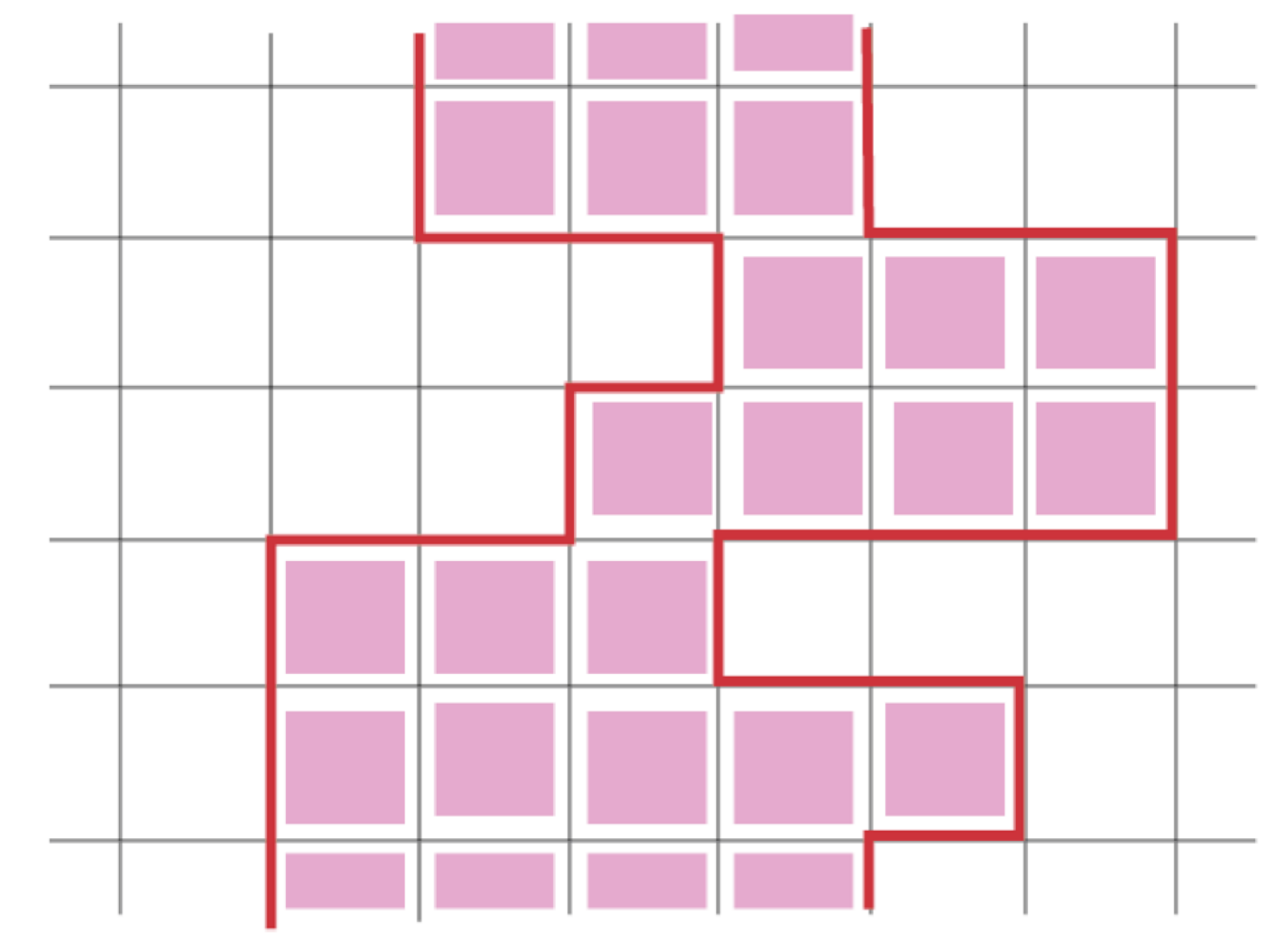}
\end{center}
\caption{Path of a heavy quark and antiquark in lattice  space-time,
  with the leading terms from the gauge theory at strong coupling.}
\label{fig:localpath}
\end{figure}

To understand the implications of this expansion, go back to
Fig.~\ref{fig:globalpath}.
In the gauge theory, each link of the path of the heavy quark and
antiquark acquires
a factor $U_{n,n+\hat\mu}$.  Each term $\tr[V_n]$ is a product of four
factors
of $U_{n,n+\hat\mu}$ arranged around an elementary square of the
lattice.   We might imagine this as a tile placed on the square.
These factors must come together to prevent the integrals over the 
gauge group from giving a zero result.  Indeed,
\beq
       \int dg   U(g)_{ij}  = 0 \quad  \mbox{while}  \quad    \int dg
       U(g)_{ij} U^\dagger(g)_{k\ell}  = c \delta_{i\ell} \delta_{kj}
       \ .
\eeq{Uintegrals}
A term in the strong-coupling perturbation theory is nonzero
only if each link has a matching number of factors of $U$ and
$U^\dagger$.
The nonzero terms correspond to tilings of the region between the
quark and antiquark paths, as shown in Fig.~\ref{fig:localpath}.

We find that the amplitude for propagation of a heavy quark-antiquark
pair is nonzero only if the entire region between the paths of these
particles is spanned by tiles.   If the quark and antiquark are far
apart, there must be gauge excitation covering every interval between
them.   This led Wilson to the conclusion that  the strong-coupling
gauge theory gives  a potential
between quarks and antiquarks of the form
\beq
       V(|\vec x_q - \vec x_{\bar q}|)  \sim   k |\vec x_q - \vec
       x_{\bar q}|
\eeq{potential}
rising linearly with distance.

For general values of the coupling, the qualitative behavior of this
potential would depend on the long-range order in  the gauge degrees of
freedom.
At strong coupling
\beq
            \VEV{ U_{n,\hat\mu}}  = 0 \ ,
\eeq{Uatstrong}
and we find the quark-confining potential \leqn{potential}.  At weak
coupling, at least in electrodynamics, there is an expansion about 
\beq
         U_{n,\hat\mu} \approx 1
\eeq{Uatweak}
that leads to  the usual Coulomb potential.   In non-Abelian
gauge theories, it is plausible that the renormalization-group flow
makes the coupling strong enough, at sufficiently large distances,
that the strong coupling region is reached and the theory is
confining.    This statement is not yet proven rigorously, but it is
supported by a wealth of numerical data~\cite{LGTreview}.

\section{Afterword}

I am ending this review almost exactly at the point where my career
intersects the story.  I came to Cornell as a graduate student in the
fall of 1973.  The first particle theory seminar that I attended was
Wilson's seminar at Cornell on the  lattice gauge theory results  that I have
just described.  A year later, the discovery of the $J/\psi$ resonance
led to striking evidence for all aspects of the picture I have
described here:   the quark model, asymptotic freedom at short
distances, a linear confining potential at large distances, even the
direct experimental verification of the number 3 in the $SU(3)$ gauge
group~\cite{quarkoniareview}.  The rout of QCD was on.

By the time I finished graduate school, QCD was already an established
theory.
I became interested in what I felt would be the next problem ripe for
solution, the physics of the spontaneous breaking of symmetry
responsible for the properties of the subnuclear weak interactions.
Thirty-five years later, that problem is still an open one, although
the recent discovery of the Higgs boson at the Large Hadron
Collider~\cite{ATLAS,CMS} surely provides an important piece of the
puzzle.

Every student of physics seeks to emulate his or her thesis advisor.
Having Ken Wilson as an advisor sets a very high standard.

The experience of working with Ken instilled some values that
continue to guide my approach to physics.  First, even when
approaching the fundamental equations of nature, a physicist should dismiss mysticism.
The universe is essentially mechanical.  There is a Hamiltonian; solve
it.
For better or worse, I find the current Standard Model of particle
physics too lacking in explanatory power, and too lacking in specific
mechanisms that might explain the fact and the consequences of its 
spontaneous symmetry breaking.

 Second, a
physicist should have a vision, and pursue it to the end.
We are not all as blessed with genius as Ken Wilson, but the mountains
are there nevertheless. Ken always climbed straight up.

\Acknowledgements

I am grateful to Belal Baaquie, Edouard Br\'ezin and John Cardy for
encouragement to write 
this review of Wilson's work, and to Jeevak Parpia and Peter Lepage
for the organization of a symposium devoted to Ken Wilson at
which this material was  first presented.   As to the content of the 
paper, I owe much to all of my teachers and colleagues at Cornell in
the 1970's. 
This work was supported by the U.S. Department 
of Energy under contract DE--AC02--76SF00515.

\end{document}